\documentclass[aps,prl,10pt]{revtex4}%
\usepackage{amsfonts}
\usepackage[T2A]{fontenc}
\usepackage[cp1251]{inputenc}
\usepackage{amsmath}
\usepackage{amssymb}
\usepackage[english]{babel}
\usepackage{graphicx}

\begin{document}

\textbf{Comment on \textquotedblleft Formation Mechanism and
Low-Temperature Instability of Exciton Rings\textquotedblright}

\bigskip

In the Letter \cite{1} the diffusive transport model (DTM) has
been used to explain the formation of a large ring around the
laser excitation spot in the spatial pattern of the interwell
exciton photoluminescence (IEPL) from the GaAs/AlGaAs double
quantum wells (DQW) \cite{2}. The calculation in \cite{1} contains
a fault that drastically affects the physical meaning of the
analytical result (formula (3) in \cite{1}). The correct result
\cite{4} is inconsistent with the experiment explanation suggested
in \cite{1}. We show that neither the diffusion model itself nor
the assumptions used in \cite{1} (and in \cite{4}) are not
applicable to the experiments \cite{1},\cite{2} for the derivation
of the ring radius.

The qualitative picture suggested in \cite{1} is as follows. Due
to the static voltage applied to the DQW structure, the potential
wells are biased that would provide the spatial separation of
electrons in one well and holes\ in another. In the absence of
laser excitation the conduction bands of the quantum wells contain
electrons due to the distributed spurious current across the DQW
structure \cite{1}. The contribution of the photogenerated
electrons to the total electron density in the wells at stationary
laser pumping is neglected. However, the same amount of the
photogenerated holes give rise to the depletion of the total
electron density (due to the interwell exciton formation) near the
laser excitation spot. It causes the diffusion of electrons from
the periphery to the spot region. The diffusion of holes goes in
the opposite direction. As a result, the hole density falls from
the center spot to the periphery and the electron one does vice
versa; at their maximal overlapping region the luminescence ring
appears. The ring radius $R$ is derived within the DTM as
\begin{equation}
R=\lambda\exp\left(  -2\pi D_{e}n_{0}/P_{ex}\right)  , \label{1}%
\end{equation}
where $D_{e}$ is the diffusion coefficient for electrons, $n_{0}$
is the equilibrium two-dimensional (2D) electron density, $P_{ex}$
is the stationary photoexcitation power, and the electron
depletion length $\lambda\gg R$ \cite{1}. According to the model
\cite{1}, the diffusion of holes is the only reason why they move
out of the laser excitation spot. However, according Eq.(\ref{1})
the ring radius does not depend on the hole diffusion coefficient
$D_{h}$ (in \cite{1} $D_{h}$ is in the exponent also), i.e. the
ring forms even if all the holes are left into the excitation spot
($D_{h}=0$). (This is because of the neglect of the photogenerated
electrons: if these are included in the model \cite{1}, due to the
exciton formation term there will be no ring at all.) In \cite{4}
it is shown in general case (i.e. regardless to $\lambda\gg R$)
that the dependence $R\left( D_{h}\right)$ appears in the DTM only
if one introduces a hole tunneling out of the well there. However,
the tunneling may be a probable process for electrons but
not for the heavy holes with the effective mass ratio $m_{h}^{\ast}%
/m_{e}^{\ast}\approx7$ in GaAs.

There are some other principal discrepancies between the
experimental data and the diffusive model results
\cite{1},\cite{4}. (1) The DTM is incapable to explain nearly the
absence of a spatially uniform IEPL between the excitation spot
region and the external ring \cite{2}. It always predicts
noticeable constant luminescence there \cite{1},\cite{4}. (2) The
model can not explain the condition on the excitation power: the
external ring appears only when the power exceeds some critical
value, $P_{ex}>\left(  P_{ex}\right) _{c}\approx250$ $\mu W$
\cite{2}. Formally, the model \cite{1},\cite{4} is purely 2D but
no ring pattern was observed for the system in the intrawell
exciton luminescence \cite{1},\cite{2}. (Note that another
experimental group \cite{61} reported that they had also observed
the ring in the case of intrawell excitons for DQW and for a
single quantum well.)

The assumptions embodied in the model \cite{1} also contain some
physical shortcomings. (1) If the conduction bands of the DQW have
essential electron occupancy in the absence of laser pumping, then
the wells cannot be considered as insulating layers and the
transverse electric field made by the static gate voltage $V_{g}$
depends on the coordinate $z$ perpendicular to the layers. The
latter contradicts the statement \cite{6} that the DQW potential
energy profile along $z$-axis has only a constant tilt due to
$V_{g}$. (2) The diffusion coefficients of the carriers seem too
small to give the diffusion length of about $100$ $\mu m$. In
particular, they can be estimated by the Einstein relation
$D_{e,h}=\left(  \mu_{e,h}/e\right)  T$, where $T$ is the lattice
temperature. The typical experimental value for the electron
mobility in a single GaAs quantum well at $T\sim1K$ is
$\mu_{e}\sim10^{6}$ $cm^{2}/\left(  Vs\right)  $ \cite{7}. The
heavy hole mobility $\mu_{h}$ is at least one order of magnitude
smaller than $\mu_{e}$. At $T\sim1K$ the values $D_{e}\sim10^{2}$
$cm^{2}/s$, $D_{h}\sim10$ $cm^{2}/s$ (cp. \cite{71}) are too small
to provide the ring radius $R\sim\sqrt{D\tau_{X}}\sim10^{-2}$
$cm$, where
$D=\min\left(  D_{e},D_{h}\right)  $ and $\tau_{X}\sim\left(  10^{-8}%
\div10^{-7}\right)  $ $s$ is the typical interwell exciton
lifetime \cite{1}. (The minimal suitable value $D\sim10^{3}$
$cm^{2}/s$ can be obtained if one takes in the Einstein relation
the lattice temperature $T\sim100$ $K$.) (3) Finally, in the model
\cite{1} electroneutrality has been violated on the macroscopic
length scale $\sim R\sim0.1$ $mm$.

Concluding the criticism, one may say that the model \cite{1} is unable to
give a consistent explanation of the experimental results \cite{1},\cite{2}.

It is advisable to explain the ring formation only by kinetics of
the photogenerated carriers, i.e. without background equilibrium
electrons and the transverse spurious current as well. The
suggestion is that at high photoexcitation power the in-plane
electric fields appear in the excitation spot region. In
particular, if the carrier densities in the spot become high
enough, the repulsive intralayer Coulomb forces between particles
become stronger than the attractive interlayer force. In this case
the intralayer elecrtic fields appear to eject "surplus" electrons
and holes from the excitation spot. Due to high mobilities of the
carriers in GaAs quantum wells at low temperature, it results in
high initial velocities of the "surplus" particles (regime of hot
carriers). In the regime the value of carrier velocity is
restricted by optical phonon emission. The estimate for the
maximal initial velocity through
$m^{\ast}v_{\max}^{2}/2=\hbar\omega_{LO}$ , where $\hbar\omega
_{LO}\approx37$ $meV$ is the optical phonon energy in GaAs and $m^{\ast}%
\sim0.1m_{e}$, results in $v_{\max}\sim10^{7}$ cm/s, a typical
saturation value for the carrier drift velocity in GaAs with the
increase of applied electric field \cite{14}. To form an exciton
the electrons and holes should cool down by emitting acoustic
phonons. At typical electron-acoustic phonon scattering time
$\tau_{e-ac}\sim10^{-9}$ $s$ \cite{17} one comes to the promising
estimate $R\sim v_{\max}\tau_{e-ac}\sim10^{-2}$ $cm$ (cp.
\cite{16}). However, a quantitative theory describing the in-plane
expansion of the hot spatially separated electrons and holes,
which finally form excitons, is absent.

Apparently, the ambipolar (or drift-) diffusion regime (see e.g.
\cite{15}) is valid for slow carriers at in-plane distance $r\sim
R$, where the ambipolar electric field $\mathbf{E}$ might play an
important role in the formation of a sharp width of the ring
(interestingly, the FWHM of the ring intensity is practically
independent on the ring radius $R$ at high $P_{ex}$ \cite{2}). The
continuity equations in this regime are given by
\begin{align}
\dot{n}_{e(h)}+\operatorname{div}\mathbf{i}_{e(h)} &
=g_{e(h)}-\Gamma,\label{2}\\
\dot{n}_{X}+\operatorname{div}\mathbf{i}_{X} &  =\Gamma-n_{X}/\tau
_{X}.\label{3}%
\end{align}
Here $n_{e}$, $n_{h}$ and
$\mathbf{i}_{e}=-n_{e}\mu_{e}\mathbf{E}-D_{e}\nabla n_{e}$,
$\mathbf{i}_{h}=n_{h}\mu_{h}\mathbf{E-}D_{h}\nabla n_{h}$ are 2D
densities and particle flux densities of free, uncoupled electrons
in plane $z=d/2$ and holes in plane $z=-d/2$, $\mu_{e(h)}$ is
electron (hole) mobility that is assumed to be a constant. The
particle flux density for excitons
$\mathbf{i}_{X}\approx-D_{X}\nabla n_{X}$, $n_{X}$ is the
interwell exciton density. The contribution from the dipole-dipole
interaction between the excitons is omitted in $\mathbf{i}_{X}$
since it appears as an above-linear correction on $n_{X}$. The
carrier generation rates $g_{e(h)}\left(\mathbf{r},t\right)$ are a
priori given functions. The general form of the exciton formation
rate reads (hereafter inessential constant prefactors are dropped)
\begin{equation}
\Gamma\left(  \mathbf{r},t\right)  =%
{\displaystyle\int}
w\left(  \left\vert \mathbf{v}_{1}-\mathbf{v}_{2}\right\vert \right)
f_{e}\left(  \mathbf{r},\mathbf{v}_{1},t\right)  f_{h}\left(  \mathbf{r}%
,\mathbf{v}_{2},t\right)  d^{2}\mathbf{v}_{1}d^{2}\mathbf{v}_{2}, \label{4}%
\end{equation}
where $f_{_{e(h)}}\left(  \mathbf{r},\mathbf{v},t\right)  $ is the
electron (hole) distribution function, so that $n_{e(h)}\left(  r,t\right)  =%
{\displaystyle\int} f_{_{e(h)}}\left(  r,\mathbf{v},t\right)
d^{2}\mathbf{v}$, and $w\left(v\right)$ is the specific exciton
formation rate. Finally, the Poisson equation for the electric
field reads (time dependence is dropped; $\varepsilon$ is
dielectric constant)
\begin{equation}
\operatorname{div}\left(  \varepsilon\mathbf{E}(r,z)\right)  =4\pi e\left[
\left(  n_{h}(r)+n_{X}(r)\right)  \delta(z+d/2)-\left(  n_{e}(r)+n_{X}%
(r)\right)  \delta(z-d/2)\right]  . \label{5}%
\end{equation}
It includes the contribution of the interwell exciton dipole fields and keeps
the electroneutrality for the free carrier system when the exciton formation
is suppressed ($n_{X}(r)=0$).

At $r\sim R$ one may put $w\left(  v\right)  \approx w_{\max}$,
then $\Gamma\left(  r\right)  \approx w_{\max}n_{e}\left( r\right)
n_{h}\left( r\right)  $ and the Eqs.(\ref{2}),(\ref{3}),(\ref{5})
with $g_{e(h)}=0$ become a closed system.

The author thanks A. S. Alexandrov, Yu. M. Kagan, F. V. Kusmartsev, L. A.
Maksimov, and S. E. Savel'ev for helpful discussions.

\bigskip

A. V. Paraskevov

\bigskip

Department of Physics, Loughborough University, Loughborough LE11 3TU, United Kingdom

Kurchatov Institute, Moscow 123182, Russia

\end{document}